\documentclass[12pt]{article}
\usepackage{amssymb}
\usepackage{amsfonts}
\usepackage{epsfig}
\usepackage{graphicx}
\usepackage{epsfig}
\usepackage{xcolor}

\begin{document}

\title{Big Missing Data: are scientific memes inherited differently from gendered authorship?}
\author{{\small Tanya Ara\'{u}jo$^{a,b}$ and Elsa Fontainha$^{a}$}\\
 {\small $^{a}$ISEG (Lisbon School of Economics \& Management) Universidade de Lisboa}  \and
{\small Rua do Quelhas, 6 1200-781 Lisboa Portugal} \and {\small $^{b}$ Research
Unit on Complexity and Economics (UECE)}}
\date{}

\maketitle

\begin{abstract}
{\small This paper seeks to build upon the previous literature on gender aspects in research collaboration and knowledge diffusion. Our approach adds the meme inheritance notion to traditional citation analysis, as we investigate if scientific memes are inherited differently from gendered authorship. Since authors of scientific papers inherit knowledge from their cited authors, once authorship is gendered we are able to characterize the inheritance process with respect to the frequencies of memes and their propagation scores depending on the gender of the authors. By applying methodologies that enable the gender disambiguation of authors, big missing data on the gender of citing and cited authors is dealt with. Our empirically based approach allows for investigating the combined effect of meme inheritance and gendered transmission. Results show that scientific memes do not spread differently from either male or female cited authors. Likewise, the memes that we analyse were not found to propagate more easily via male or female inheritance.}

\end{abstract}



\section{Introduction}

Researchers now live in an era of a New Data Frontier, a term coined by Maryann Feldman and co-authors in a recent paper \cite{Feldmana}. Very large databases available in digital format contribute to enlarge the horizons of knowledge in multiple domains, such as social media communication, health, business, finance, economics, invention and innovation, and scientific diffusion and progress. Some recent literature based on big data spans multiple data sources, disciplines and applications, including \emph{i)} financial markets forecasting through text-based information from news and social media \cite{Bukovina};\emph{ ii)} investor sentiment research built on machine learning and natural language processing \cite{Curme}; \emph{iii)} analysis of co-occurring terms in opinion dynamics \cite{Ban12}; \cite{Weis} and \emph{iv)} innovation diffusion based on epidemic models \cite{Pulkki}.

In addition, various datasets have been used as sources for the identification of patterns of knowledge diffusion and scientific progress at both individual and institutional levels. Big data on scientific knowledge, in particular, has received considerable attention. Studies have explored both traditional scientific data sources of publications and patents \cite{Feldmanb}, as well as, new data sources like University web-sites \cite{Geuna} or communication within large virtual academic communities \cite{Fontainha}.

Notwithstanding, most analyses of knowledge transmission, channels and mechanisms are based on classical large scientific databases like Google Scholar, PubMed, Scopus and Web of Science, which have been intensely compared, discussed and evaluated (\cite{Adriaanse}, \cite{Bakka}, \cite{Falagas}, \cite{Har}, \cite{Kul}, \cite{Meho}).

Besides providing raw data on publications and patents, some large scientific databases publish science metrics, such as Impact Factor per journal, h-index per author and citation scores. Studies using citation indicators have provided insights into scientific performance and impact \cite{Evans}, and knowledge dissemination among individuals \cite{Sorenson}, firms \cite{Aha} or regions \cite{Bornmann}, as well as, between universities and firms \cite{Azagra}. Citation content and frequency also feed network approaches used in the study of social contagion mechanisms. Research collaboration and co-authorship help to discover patterns of collaboration within scientific communities of authors, inventors or innovators \cite{Ara17}.

Tobias Kuhn and co-authors \cite{Kuhn} study the spread of scientific knowledge using Dawkins' concept of meme, the cultural analogy of gene in the context of  genetic evolution. Illustrations of a scientific meme as a replicator are provided in The Selfish Gene book by Richard Dawkins: "If a scientist hears, or reads about, a good idea, he passes it on to his colleagues and students. He mentions it in his articles and his lectures. If the idea catches on, it can be said to propagate itself, spreading from brain to brain. If the meme is a scientific idea, its spread will depend on how acceptable it is to the population of individual scientists; a rough measure of its survival value could be obtained by counting the number of times it is referred to in successive years in scientific journals." \cite{Daw}.

Although the idea of memes is not completely original, as Dawkins acknowledges, it has received growing interest ever since. The concept of meme has been explored in several scientific areas. In Economics, Robert Shiller recently drew attention to memes and narratives in his Presidential Address delivered at the American Economic Association meeting: "There is a daunting amount in the scholarly literature about narratives, in a number of academic departments, and associated concepts of memetics, norms, social epidemics, contagion of ideas. While we may never be able to explain why some narratives go viral and significantly influence thinking while other narratives do not [$...$] We economists should not just throw up our hands and decide to ignore this vast literature." \cite{Shiller}.

 Here, we aim to investigate if scientific memes are inherited differently from gendered authorship. Since authors of scientific papers inherit knowledge from their cited authors, once authorship is gendered (by applying methodologies that enable the gender disambiguation of authors), we are able to characterize the inheritance process with respect to the frequencies of scientific memes and their propagation scores depending on the gender of the authors. Would female inheritance - represented by the citations of female authors - favor the propagation of some specific meme? Likewise, would some particular memes propagate more via male inheritance?

Moreover, our paper seeks to build upon the previous literature about gender aspects in research communication, collaboration and co-authorship (\cite{Ara17},  \cite{Aste}, \cite{Bozeman}, \cite{Brooks}, \cite{Gonzalez}, \cite{Sugimoto}, \cite{Tartari}, \cite{Van}, \cite{Viana},  \cite{Ynalvez}) and scientific outcome impact by gender (\cite{Beaudry}, \cite{Cope}, \cite{Melo}, \cite{Friet}, \cite{Giu}, \cite{Ghi}, \cite{Hunt}, \cite{Jung}, \cite{Mau}, \cite{Meng}, \cite{Miha}, \cite{Okon}).

By including gender in the study of knowledge spread and adding a gender perspective to the \emph{'spreading of good ideas, from brain to brain'}, to adopt
Dawkins's words, our empirically based research aims to contribute in three ways to the improvement of the understanding of the way knowledge spreads:
\begin{itemize}
 \item it sheds some light on previous mixed and puzzling results about women in science;
 \item it adds the meme inheritance notion to traditional citation network analysis and
 \item it accurately identifies the gender of authors, dealing with the issue of missing data in large scientific databases.
\end{itemize}
The remainder of this paper is structured as follows: Section 2 briefly describes some scientific databases calling attention to the lack of information about the gender of citing and cited authors. Section 3 presents the data and the methodologies used in the paper. Section 4 presents and discusses the results from the empirical analyzes. In the final section conclusions, policy implications and some promising research avenues are provided.

\section{Big Missing Data}

Four databases (Web of Science WoS, SCOPUS, Google Scholar GS and PubMed), and three repositories (arXiv, RepEc and BASE-Bielefeld) were searched in order to explore the possibilities of extracting information regarding the gender of both citing and cited authors and the memes found in the abstracts of citing and cited records (papers).
The following criteria were adopted to select the datasets to examine in detail: size and accuracy of the data; tracking citation possibilities; down-loading capabilities; and possibility to gather, for each record, at least, its title and abstract.

Several studies have compared the coverage, features, and citation analysis capabilities of GS, PubMed, SCOPUS and WoS. These comparative studies usually focus on a particular research topic like biomedical information \cite{Falagas}, medical journals \cite{Kul}, oncology and condensed matter physics \cite{Bakka} library and information science \cite{Meho} or environmental sciences \cite{Adriaanse}. Other studies address only the accuracy of one database \cite{France}. This literature, however, fails to systematically review the citation analysis linked with the author full identification.

\bigskip

\textbf{Web of Science (WoS)}

\bigskip

The Web of Science (WoS), formerly the ISI Web of Knowledge, is self defined as the "gold standard for research discovery and analytics" \cite{wos} and the primary research platform for information in the sciences, social sciences, arts, and humanities. It uses cited reference search to track past research and screen current advances in over 100 years worth of content that is fully indexed, including 59 million records and backfiles dating back to 1898. Web of Science consists of six databases containing information gathered from thousands of scholarly journals, books, book series and other scientific outcomes. The Master Journal List includes 22,832 titles. The databases included in WoS are: Science Citation Index Expanded (SCI-Expanded); Social Sciences Citation Index (SSCI); Arts \& Humanities Citation Index (A\&HCI); Conference Proceedings Citation Index - Science (CPCI-S); Conference Proceedings Citation Index - Social Sciences \& Humanities (CPCI-SSH); and Emerging Sources Citation Index (ESCI). The WoS also includes two chemistry databases: Index Chemicus (IC) and Current Chemical Reactions (CCR-Expanded).

The Web of Science adopts a selection process for the inclusion of journals in its content coverage \cite{Testa}. The most frequent criticisms to WoS are the bias to American-based, English-language journals, failure to completely cover other citation sources (e.g. books) and failure to include citations out of the WoS database. Despite the criticism, WoS is often used worldwide in scientometrics analysis based on information articles and articles citation on a subject (\cite{Adriaanse}, \cite{Har}, \cite{Sugimoto}, \cite{Lariviere}). The WoS has features for browsing, searching, sorting, saving and exporting data. A citation report can be generated by author (or by institution, etc.), and a citation map can be produced. Each record (an article) can be graphically represented or mapped,  linking the record to all the records that cite or are cited by the target record. Most of the articles comprise an abstract and a set of keywords. The number of keywords varies across journals and scientific domains.
From 2006 onwards the reporting of author's name in WoS changed with the inclusion of the full name (given and family name). However, the full name of cited authors, i.e., the authors in the bibliographic list of references of each article is not provided.

\bigskip

\textbf{SCOPUS}\bigskip

SCOPUS, developed and owned by Elsevier, is presented on its own Web page as "the largest abstract and citation database of peer-reviewed literature: scientific journals, books and conference proceedings" \cite{scopus}. It includes 66 million of records, 22,748 peer-reviewed journals and 7,7 million conference papers. SCOPUS includes publications from Sciences, Social Sciences and Art and Humanities. SCOPUS both covers more journals than WoS and provides better coverage of the non-North American sources. Most of the articles comprise abstracts. The citations of an author and the articles that cite the original article (using Citation Tracker) make it possible to base the analysis of citations  on different criteria and enable the researcher to create an exportable spreadsheet of the citations, which may or may not include self citations. Author names in SCOPUS can be arranged differently. Consequently, there are an unknown share of authors in the database where the given name is missing. Thus, the SCOPUS database does not allow for the gender disambiguation of authors in the bibliographic list of references of the articles. In a  recent publication prepared by Elsevier, the SCOPUS data are combined with other data sources in order to obtain information on the first names and gender of the authors \cite{Elsevier}.

\bigskip

\textbf{Google Scholar (GS)}\bigskip

The Google Scholar was created in 2004 and comprises all fields of knowledge and several types of documents, including abstracts, peer-reviewed and non-peer-reviewed papers, print and electronic journals, conference proceedings, books, theses, dissertations, preprints papers, technical reports, monographs, conference proceedings, patents, and legal documents. GS neither  defines the number of journals covered nor the time span of the database. Thus, for a document with the same title and authorship, it includes all the versions available online. Because the content coverage is unknown, there is consensus among the scientific community that it must be used with caution and is not suitable to analyze citations because of the inclusion of several versions of the same paper. Authors can create a Google Scholar Account profile. The allocation of articles to authors is carried out automatically and frequently some scientific outcomes are wrongly  attributed.

\bigskip

\textbf{Pub Med}\bigskip

PubMed is an important resource for clinicians and researchers. The de- veloper/owner is the National Center for Biotechnology Information (NCBI), US National Library of Medicine (NLM) National Institutes of Health. The dataset includes Medline (1966-present), old Medline (1950-1965), PubMed Central, and other NLM databases. Citations are not provided. Instead, for each article, there is information about similar articles. Presentation of the name(s) of the authors is incomplete; as it does not include their given names.

\bigskip

\textbf{RePEc Repository}\bigskip

The Research Papers in Economics (RePEc) is developed by volunteers from 89 countries to promote the dissemination of research in Economics and associated fields. It includes working papers, journal articles, books, book chapters and software components in a total of 2 million research outputs from 2,300 journals and 4,300 working paper series. There are 48,000 authors registered, and they are ranked according to the citations received by their scientific outcomes.
The citations coverage of RePEc is in general incomplete compared with WoS and Scopus. The citations in RePEc are collected by an experimental project, CitEc and only a minority of all works can be analyzed.  Within RePEc, a Genealogy project is being constructed in a voluntary base to create a dataset of advisors and advisees \cite{repec}.

\bigskip

\textbf{BASE Bielefeld Academic Search Engines}\bigskip

BASE, one of the largest search engines for academic resources, indexes more than 100 million documents (about 60\% full text Open Access) from more than 5,000 sources. It contains different kind of documents: text, image-video, software, and datasets). The total of 73,595,901 text documents comprises, among others, books and book parts, article contributions to journals/newspapers, patents and theses.

\newpage

\textbf{ArXiv}

\bigskip

ArXiv is a pre-print archive of working papers in Physics, Mathematics, Computer Science, Quantitative Biology, Quantitative Finance and Statistics. The repository arXiv, where High Energy Physics belongs, was founded in 1991 by Paul Ginsparg, a theoretical physicist at Cornell University, and since then it has received growing interest and use from the scientific community \cite{Gin}. A study comparing this pre-publication repository with publication databases has been carried out by Bar-Ilan \cite{Bar}.

There are several studies about the research impact of the material in the repository of arXiv namely using citation analysis (\cite{Brody}, \cite{Davis}, \cite{Evans}, \cite{Goldberg}, \cite{Haque}). Other studies use arXiv to identify trends and build the agenda for future research in multiple scientific domains (\cite{Haque}, \cite{Lariviere}).

\bigskip

\textbf{From name to gender}

\bigskip

In fact, the two major bibliographic databases, Web of Science (WoS) and Scopus, both of which cover several scientific domains and many types of scientific outputs, do not include the information needed to answer our research questions. The large databases and repositories with scientific outputs, as well as, most of the repositories, independently of the coverage (by domain, period or type of document) and citation search strategies, produce quantitatively and qualitatively different citation material.

Given the goal of this paper, the databases described have strong weaknesses resulting from the absence of gender information about the authors. The large majority of bibliometric and patent databases do not include information about the gender of the author, inventor or innovator. Under certain conditions this information can be obtained indirectly through the given or family name of the author. The situation is worse with regard to the transmission of knowledge (from citing to cited author), because the databases provide neither information on citations by gender nor the full name of the cited author, i.e., authors in the bibliographic list of references. Thus, given this lack of information, the only way to overcome these weaknesses is to obtain the information from the first name or the family name of the contributors concerned.

It is possible to generate the missing information from the given name of the author if it exists, which is rare. This procedure was adopted here to deal with missing data on gender. In brief, some databases include the full name of the authors, which enables, at least partially, the identification of his/her gender. However, the bibliographic list of references (citations) in each article does not provide the full name of the cited authors. This lack of information is solved in the present research by using the dataset provided by Stanford Network Analysis Platform (SNAP) together with GitHub package - \emph{Predicting Gender from Names using Historical Data} (\cite{Blevins}, \cite{Mullen}) - for the gender disambiguation of authors.

\bigskip

\textbf{SNAP}

\bigskip

Stanford Network Analysis Platform (SNAP) is a general purpose network analysis and graph mining library. Among the Stanford Large Network Dataset Collection we were able to download the dataset recorded from the repository arXiv: the hep-th High Energy Physics \cite{Lesko}. The detailed description of this dataset is presented in the next section.

\section{Data and Methodology}

We used the dataset recorded from the repository arXiv, the hep-th High Energy Physics - Theory and provided by Jure Leskovec at Stanford Large Network Dataset Collection \cite{Lesko}.
The data covers papers in the period from January 1993 to April 2003 (124 months) within a dataset of 29,555 papers and 352,807 links.

The available dataset is organized in three files, two of which were the main source of the work herein presented:

\begin{center}
\begin{tabular}{lll}

1) \texttt{cit-HepTh-abstracts} & includes: & \\

&\small{\textbullet}  \texttt{Paper} $id$  & \small{\textbullet} \texttt{Authors}\\
& \small{\textbullet}  \texttt{Title}  & \small{\textbullet}  \texttt{Abstract}\\

2) \texttt{cit-HepTh} provides: & \small{\textbullet}  list of directed & edges from \texttt{Paper} $i$ to \texttt{Paper} $j$ \\
\end{tabular}
\end{center}

Each directed link in the citation network \texttt{cit-HepTh} indicates that \texttt{paper} $i$ cites \texttt{paper} $j$. If a paper cites, or is cited by, a paper outside the dataset, the list does not contain any information about this.

 For each paper in the dataset, the field \texttt{Abstract} includes the corresponding abstract of the paper. The field \texttt{Authors} provides the full name of the authors in 70\% of the observations. Since most of the the papers in this dataset comprise the first (given) name of the authors, we were able to classify the citing and cited authors by gender.

\bigskip

\textbf{Gender disambiguation of authors}

\bigskip

While bibliometric databases have been the main sources for the study of knowledge spread, because they do not include information separated by male and female authors, the study of the scientific production by gender has been frequently  limited to surveys or case studies \cite{Friet}. When the number of observations is low, gender allocation is done manually. Automatically allocating gender to researchers depends on the availability of:
\emph{(i)} databases with gender information that allow matching by researcher name or code. For example, Abramo and coauthors \cite{Abramo} combine data from WoS with data from the Italian Ministry of Education, Universities and Research;
\emph{(ii)} databases with a list of male and female given names in different languages, as for instance, the database built under a EU project \emph{Improving Human Research Potential and the Socio-Economic Knowledge Base} (\cite{Naldi1}, \cite{Naldi2}) and the method used in a recent report from Elsevier \cite{Elsevier};
\emph{(iii)} language specific characteristics that allow for systematically allocating gender from each researcher's given name as, for example, the Portuguese given names \cite{Ara17}.  Polish names allow to allocate gender from the family name \cite{Kosmulski}\footnote {Most of the Asian researchers when publishing in English-language journals have to adopt a phonetic version of the given and family names and this creates ambiguities to the authors' gender attribution \cite{Qiu}.}.

This research adopts the methodology mentioned in \emph{(ii)} and uses GitHub (\cite{Blevins}, \cite{Mullen}) as the data source for the gender disambiguation of authors.

\subsection{Sample Characteristics}

Tables 1 and 2 display an overview of the basic information compiled from arXiv: the hep-th High Energy Physics \cite{Lesko} dataset after the gender disambiguation of authors.
Author gender was accurately assessed for 70\% of the papers. As earlier mentioned, the loss of information in the process of disambiguation is in line with other studies (\cite{Beaudry}, \cite{Lariviere}). A gendered paper is a paper that enables the gender disambiguation of at least its first author. A gendered author is an author which enables the identification of his/her gender. The average number of papers by gendered author is 2.1. A gendered link is a link between two gendered papers. 58\% of the citations are gendered links. There is an overlap of approximately $2/3$ of the papers that are both citing and cited papers in the citation network. This ratio also applies when we consider just the papers with gendered authors, as the values in Table 1 show.

\begin{center}
\begin{tabular}{l|c|c}
\hline
 & All & with Gender \\
 \hline
	\small {Number of papers}&  \small{29,555}&  \small{20,657}\\
	\small{Number of citations}& \small{352,807}&  \small{206,405} \\
	\small{Number of Citing papers} ($Ci$)& \small{25,058}&  \small{17,230} \\
    \small{Number of Cited papers} ($Ce$)&\small{ 23,180} &  \small{15,596}\\
	\small{Size of} ($Ci \bigcup Ce$)& \small{27,770} &  \small{19,153} \\
    \small{Size of} ($Ci \bigcap Ce$)& \small{20,468 } & \small{13,673}\\
\hline
\end{tabular}
\medskip

{\small Table 1: Paper-centered information from the original dataset after gender disambiguation of authors.}

\end{center}

Table 2 provides author-centered information considering just the first author of each paper, which corresponds to 9,830 unique authors. Although 44.6\% of papers have a second author (5.1\% have a third author, and just 0.009\% have a fourth author), in the current research and for simplicity reasons, only the first author of each paper was considered. Its distribution by gender yields 1~,079 female and 8,751 male authors. The percentage of female authors in the citing papers is 10.9 and in the cited papers is 9.0. We found that 22.6\% of the citations are self citations. Female and male authors display percentages of 24\% and 24.9\% of self citations, respectively.

\begin{center}
\begin{tabular}{l|c|c|c|c}
\hline
 & All & Female & Male & missing\\
 \hline
	\small{Number of 1st authors}            &\small{14,099} & \small{1,079}  & \small{8,751}  & \small{4,269}\\
    \small{Number of 2nd Authors}            &\small{6,496}  &\small{687}    & \small{4,200} & \small{1,609} \\
	\small{\% of 1st authors \textbf{citing} by gender}& 100 & \small{10.9}&  \small{89.1} & --\\
	\small{\% of 1st authors \textbf{cited} by gender}& 100 & \small{9.0}&  \small{91.0} & --\\
\hline
\end{tabular}

\medskip

{\small Table 2: Author-centered information after the disambiguation of authors.}
\end{center}

 The difference between the percentage of female cited and male cited authors seems to mirror the universe of the publications' authorship (citing authors) by gender. Because the cited papers correspond by definition to a period of time before the citing papers, and the proportion of female researchers for the subject area Physics and Astronomy has tended to increase in developed countries \cite{Elsevier}, the slight difference found (from 10.9 and 89.1 to 9.0 and 91.0) merely reflects a potential number of citable papers authored by women that is smaller than that of the citing papers.

\begin{figure}[h]
\begin{center}
\psfig{figure=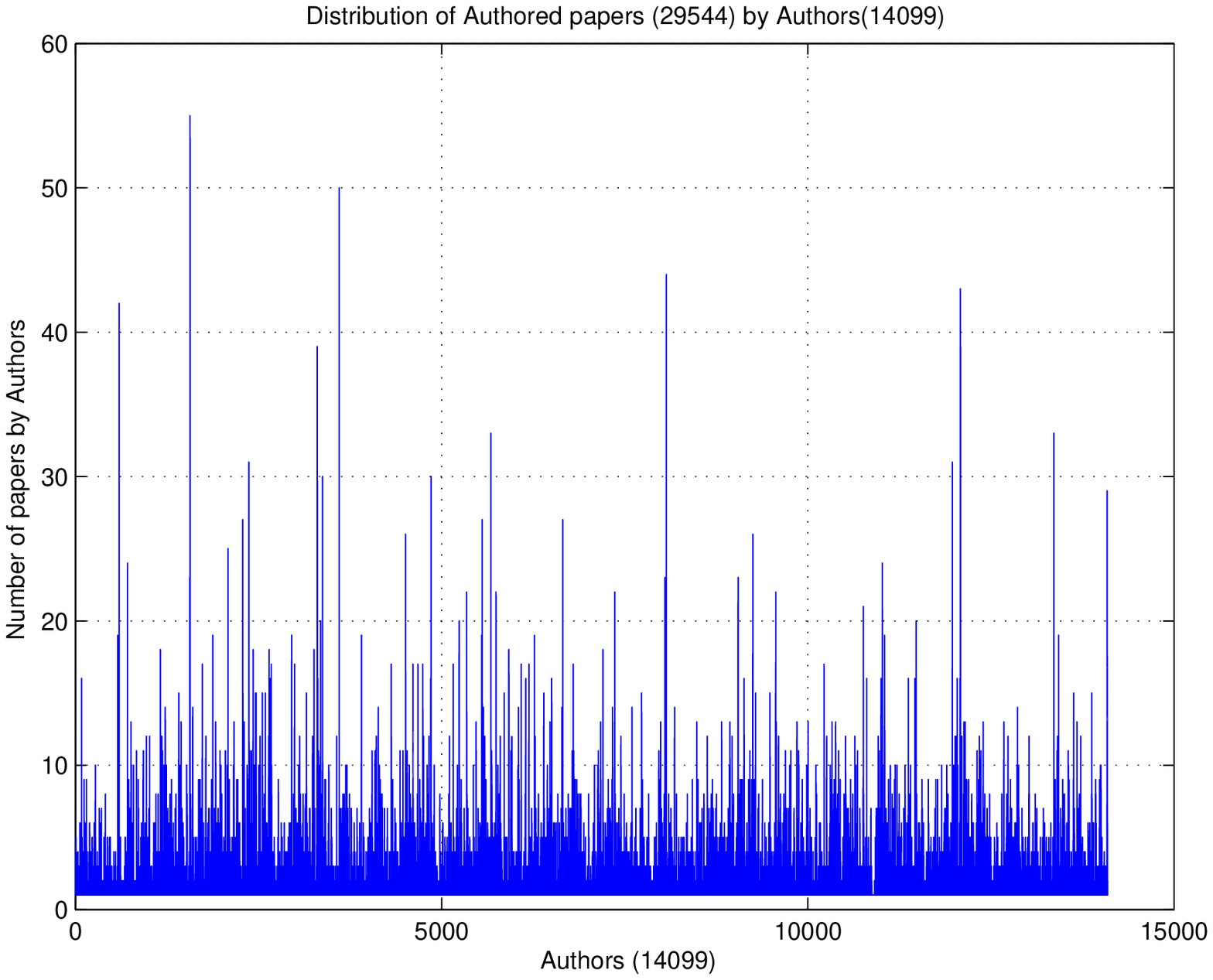,width=6truecm}\\
\psfig{figure=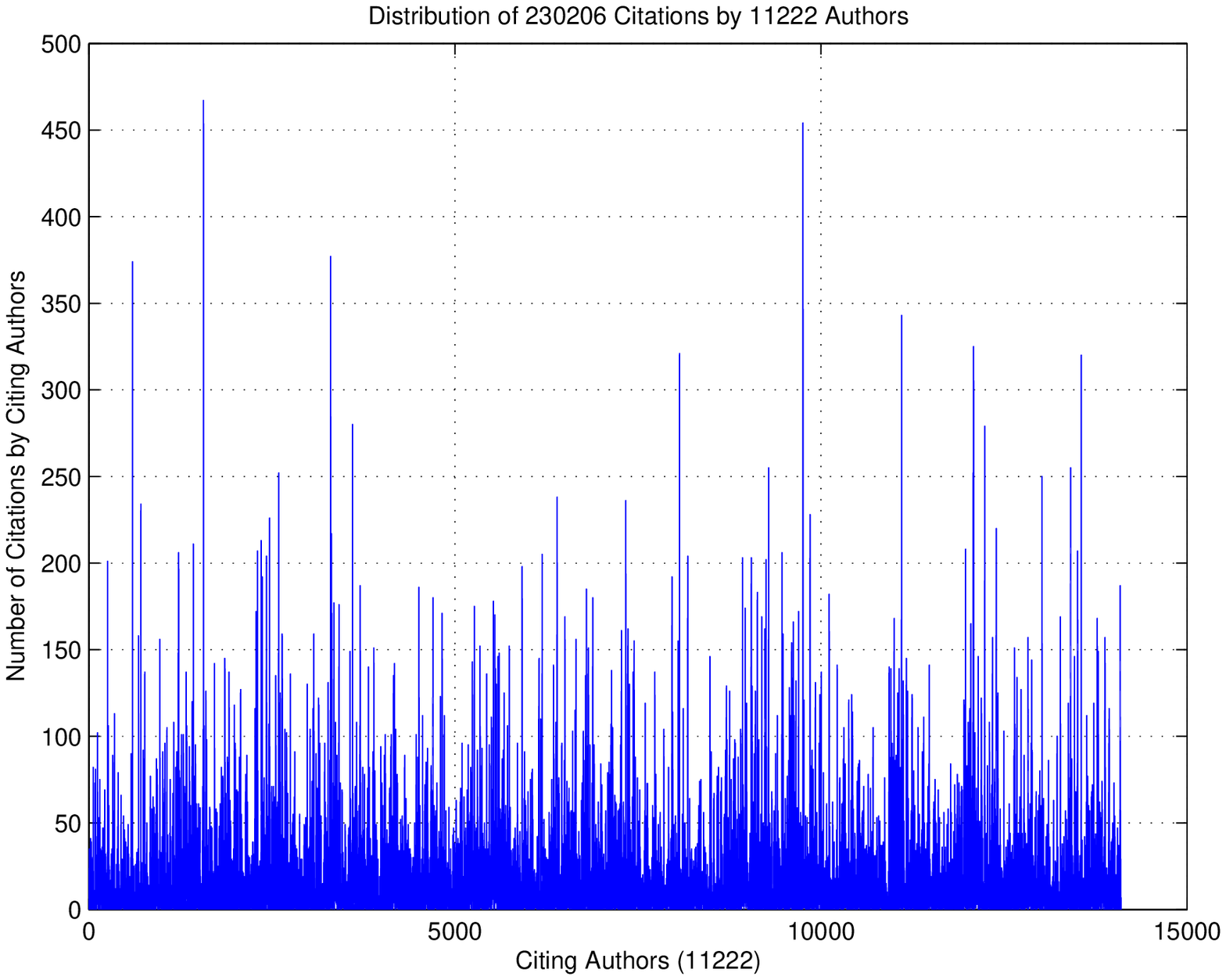,width=6truecm}
\psfig{figure=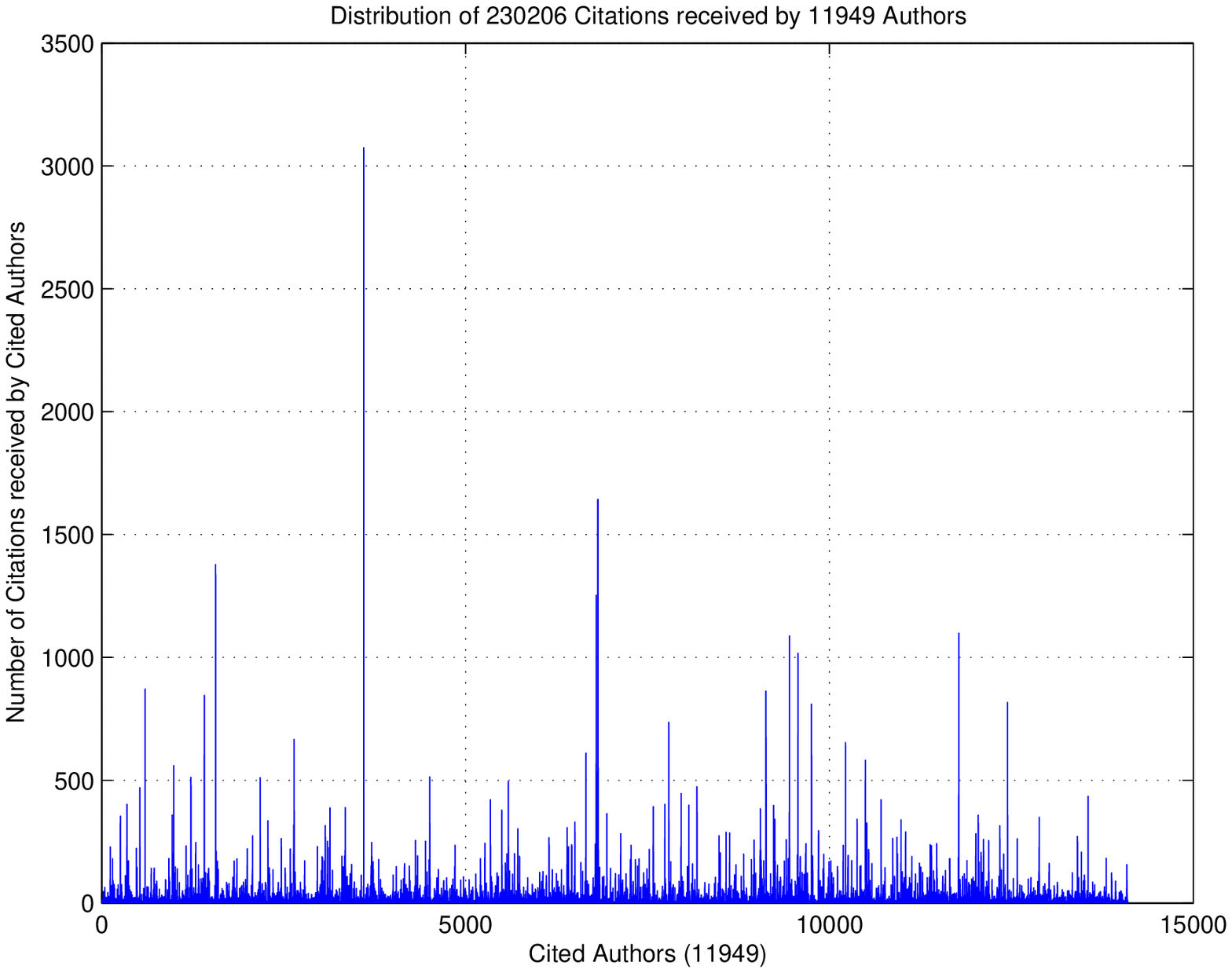,width=6truecm}
\caption{The distributions of the number of \emph{(a)} authored papers by authors, \emph{(b)} citations by citing authors and \emph{(c)} citations by cited authors.}
\end{center}
\end{figure}

\begin{figure}[h]
\begin{center}
\psfig{figure=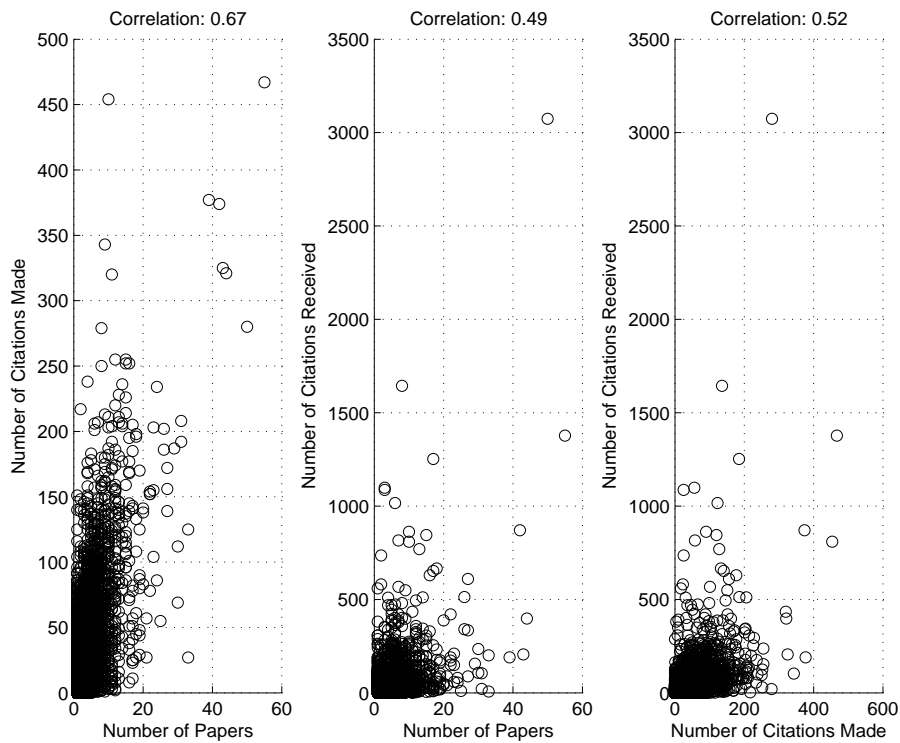,width=12truecm}
\caption{Author-based scatter plots and correlation coefficients between: \emph{(a)} the number of papers authored and the number of citations made, \emph{(b)} the number of papers authored and the number of citations received, and \emph{(c)} the number of citations made and the number of citations received.}
\end{center}
\end{figure}

Figure 1 shows the distributions of the number of papers by authors (14,099: 1,079 females, 8,751 males and 4,269 missing). It also show the distribution of the number of citations by either the citing or the cited authors in the sample. Each value in the $x$ axis represents the author (numeric) identification. Numbers were assigned according to the name of the authors ranked in alphabetical order.
Figure 1(a) shows 29,544 papers distributed by 14,099 authors (there were 11 papers without the author name, just 29,544 papers were thus considered). Figures 1(b) and 1(c) represent, respectively, the citations by citing and cited authors. For example, the highest value in Figure 2(b) shows that there is an author that cites more than 450 other authors. At the same time, the highest value in Figure 1(c) shows that there is an author that receives more than 3,000 citations.

Scales are different because the distribution of the citations made is much more homogeneous than the distribution of the citations received. In terms of a citation network and considering a direct network of authors, the average number of in-coming links per author is $20.50$ and the average number of out-going links is $19.26$. Therefore, although on average, the number of in-coming links is close to the number of out-going ones, the latter are much less equally distributed. There are authors receiving more than 1,500 and even more that 3,000 citations. A much more balanced distribution characterizes the citations made, where the most citing author does not go beyond 467 citations, as the second plot in Figure 1 shows.

The three scatter plots in Figure 2 show the strength of the correlation between three pairs of observations (the number of papers authored, the number of citation made, and  the number of citations received) that were accounted for each author in the dataset. Therefore, each mark (o) in the plots represents an author, being the $x$ and $y$ coordinates given by one of the following quantities: the number of papers authored, the number of citation made, and  the number of citations received.  Figure 2(a) shows the scatter plot of the number of papers authored against the number of citations made, Figure 2(b) shows the number of authored papers against  the number of citations received, and Figure 2(c) concerns the number of citations made against the number of citations received.

The correlation coefficients (shown in the title of each scatter plot) were computed over the entire authors set (14,099 authors). As expected, the authors who are more productive increase the possibility of citing other authors since their authored papers are the citation vehicle. The correlation coefficients show that the strongest correlation (0.67) was found between the number of papers authored and the number of citations made per author. The lowest correlation between the number of papers authored and the number of citations received reflects the lag between the publication of the scientific output and acknowledgement by peers, as well as, the relation between quantity and quality. Sometimes the more productive authors (productivity evaluated by the number of papers) are not those that are cited more often.

\subsection{Memes Selection}

Following the approach of Kuhn and co-authors \cite{Kuhn}, our research is driven by the characterization of the propagation (or inheritance) mechanism of memes and not just by their frequency of occurrence as is usual in citation analysis.

A first step into this direction is the selection of a sub set of memes among the whole set of the most frequently occurring words in the entire set of 29,555 papers. Our meme selection process starts by using the word-counting procedure of Voyant Tools (\cite{voy}, \cite{Sinclair}). Voyant Tools allows for defining a list of words to be excluded from the word-counting procedure (\emph{stopwords}). Typically, a \emph{stopword} list contains functional words that do not carry much meaning, such as determiners and prepositions ("in", "to", "from", among others).
Table 3 shows 40 memes selected among the most frequently occurring words in the abstracts (without \emph{stopwords}) and ranked by frequency of occurrence.
It also shows the frequency of each selected meme computed from both the 29,555 papers and from the 20,657 gendered papers.

The memes selected correspond to the most frequent words in our sample that carry a specific meaning in the field of High Energy Physics.  Many frequent words like "theory", "dimensional" or "field" were excluded since they are not enough specific, occurring also frequently in papers found in other scientific areas. These words were excluded together with functional words because we are interested in the thematic similarities of the papers as opposed to, for instance, stylistic similarities between different authors. Therefore, we disregarded functional words, as well as, those words common to many other scientific areas.

\begin{center}
\begin{small}
\begin{tabular}{l|l|l|l|l|l}
\hline
Rank & Meme & $F_m$ & Rank & Meme & $F_m$ \\
\hline
1 & space & 	9,249  &
2 & gauge & 	8,082\\
3 & string & 	7,517 &
4 & quantum & 	6,275\\
5 & symmetry & 	5,682 &
6 & brane & 	5,153\\
7 & mass & 	5,082 &
8 & gravity & 	4,621\\
9 & group & 	4,600 &
10 & conformal & 	3,389\\
11 & potential & 	3,331 &
12 & spin & 	2,604\\
13 & hole & 	2,395 &
14 & supersymmetry & 	2,220\\
15 & supergravity & 	2,118 &
16 & topological & 	2,079\\
17 & phase & 	2,068 &
18 & abelian & 	2,034\\
19 & magnetic & 	1,983 &
20 & manifold & 	1,967\\
21 & matter & 	1,829 &
22 & spacetime & 	1,812\\
23 & vacuum & 	1,802 &
24 & coupled & 	1,795\\
25 & tensor & 	1,763 &
26 & massless & 	1,654\\
27 & renormalization & 	1,418 &
28 & cosmological & 	1,393\\
29 & gravitational & 	1,362 &
30 & bosonic & 	1,352\\
31 & chern & 	1,277 &
32 & temperature & 	1,172\\
33 & lattice & 	1,033 &
34 & discrete & 	1,023\\
35 & fermionic & 	981 &
36 & relativistic & 	932\\
37 & superconformal & 	752 &
38 & singularity & 	727\\
39 & cohomology & 	465 &
40 & hierarchy & 	464\\
\hline
\end{tabular}
\end{small}
\medskip

{\small Table 3: The absolute frequency of 40 selected memes from the frequently occurring words in the abstracts of the 29,555 papers.}
\end{center}

\begin{figure}[h]
\begin{center}
\psfig{figure=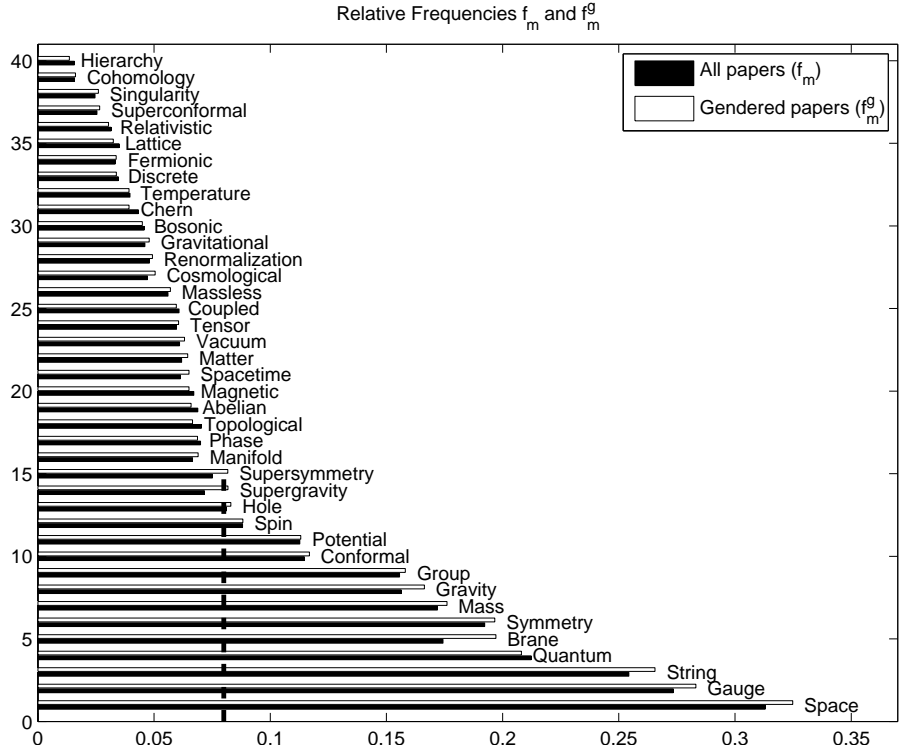,width=12truecm}
\caption{The relative frequency of occurrence of 40 memes computed from all 29,555 papers ($f_m$) and from 20,657 gendered papers ($f_m^g$).}
\end{center}
\end{figure}

Figure 3 shows the relative frequencies (the ratio of papers carrying the meme in each subset of papers) of each selected meme in Table 3. The relative frequencies are computed from both the 29,555 papers and from the subset of the 20,657 gendered papers. There are some small differences in the values of the relative frequencies depending on whether they are computed from either the 29,555 papers ($f_m$) or from the subset of the 20,657 gendered papers ($f_m^g$). Those differences are, on average, smaller than 5\% of $f_m$ and in two thirds of the memes the value of $f_m^g$ is greater than the corresponding $f_m$ value, meaning that, the relative frequencies of $2/3$ of the selected memes slightly increase when computed from the gendered papers. The vertical dashed line in Figure 3 points out the 15 memes whose frequency of occurrence in the subset of gendered papers is above $0.08$. In the next section, we compute the propagation score of these 15 memes and discuss its relation with the frequency of occurrence.


\section{Results and Discussion}

Since authors of scientific  papers inherit knowledge from their cited authors and once authorship is gendered, our research questions can be rephrased:

\begin{itemize}
\item Is the frequency and propagation of a meme (from paper cited to paper citing) influenced by the gendered cited paper?
\item Do the selected memes spread differently from either male or female cited authors?
\end{itemize}

To answer these questions we characterize the inheritance process with respect to the frequencies of memes and their propagation scores depending on the gendered authorship of the cited papers.

Departing from such a gender-oriented perspective and restricting our sample to the set of 20,657 gendered papers, two indicators are computed for each selected meme: the relative frequency and the propagation score \cite{Kuhn}.

 As already mentioned, the relative frequency of a meme computed from the set of (20,657) gendered papers ($f_m^g$) is the ratio of papers carrying the meme in this subset. The propagation score $P_m^g$ is given by:

\begin{equation}\label{1}
    P_m^g=\frac{d_{m\rightarrow m}/d_{\rightarrow m}}{d_{m\nrightarrow m}/d_{\nrightarrow m}}
\end{equation}

where $d_{m\rightarrow m}$ is the number of papers that carry the meme $m$  and cite at least one paper carrying this meme, while $d_m$ is the number of all papers (meme carrying or not) that cite at least one paper that carries the meme $m$. Following Kuhn and co-authors \cite{Kuhn}, we also compute $d_{m\nrightarrow m}$ as the number of papers that carry the meme $m$  and cite at least one paper carrying this meme, and $d_\nrightarrow m$ is the number of all papers (meme carrying or not) that do not cite a paper that carries the meme $m$.

Since in $P_m^g$, $g$ stands for gendered, its computation is made from the citation network of (206,405) gendered links.
When computing the propagation score for each specific gender ($P_m^F$ and $P_m^M$), we constrain the subsets of links being considered so that the cited papers conform to each specific gender. Therefore, in computing the female (male) propagation score of a meme $P_m^F$ ($P_m^M$), the terms $d_{m\rightarrow m}$, $d_{\rightarrow m}$,$d_{m\nrightarrow m}$ and $d_{\nrightarrow m}$ account just for the cited papers of a female (male) author.

\begin{figure}[h]
\begin{center}
\psfig{figure=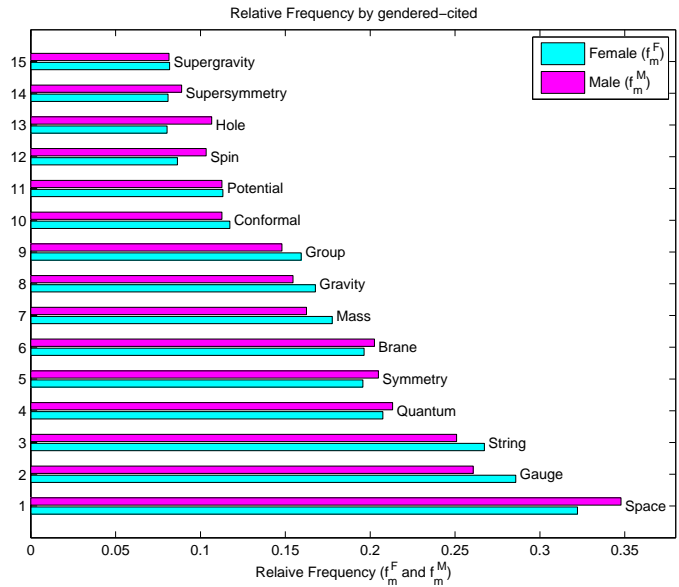,width=9truecm}
\caption{The relative frequency of memes ($f_m^{F}$ and $f_m^{M}$) in gendered papers.}
\end{center}
\end{figure}

Figure 4 and 5 show, respectively, the values of the relative frequencies ($f_m^{F}$ and $f_m^{M}$) and propagation scores ($P_m^{F}$ and $P_m^{M}$) of the 15 memes whose frequency of occurrence in the subset of gendered papers is above $0.08$. The only noteworthy difference in the propagation score by gender concerns the value obtained for the meme "Spin". In this specific case, the propagation score via male inheritance is stronger than via female inheritance. In the other 14 cases, results confirm the almost absence of any difference between female and male transmission of memes.

\begin{figure}[h]
\begin{center}
\psfig{figure=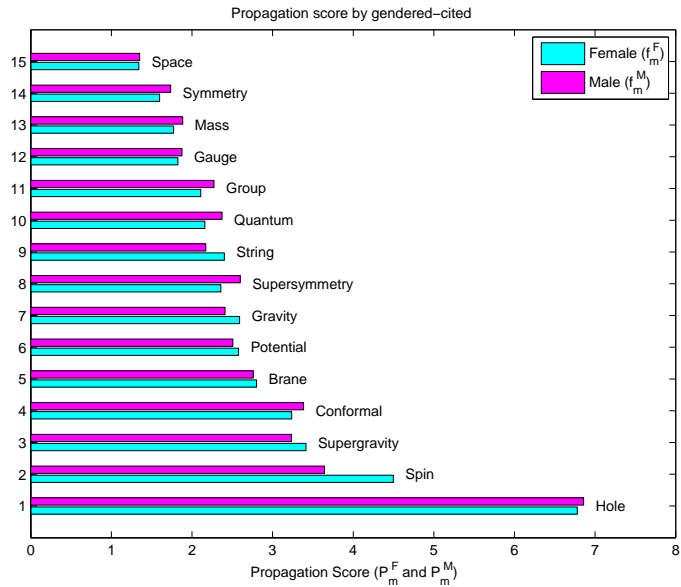,width=9truecm}
\caption{The propagation score of memes ($f_m^{F}$ and $f_m^{M}$) by gendered cited.}
\end{center}
\end{figure}

\begin{figure}[h]
\begin{center}
\psfig{figure=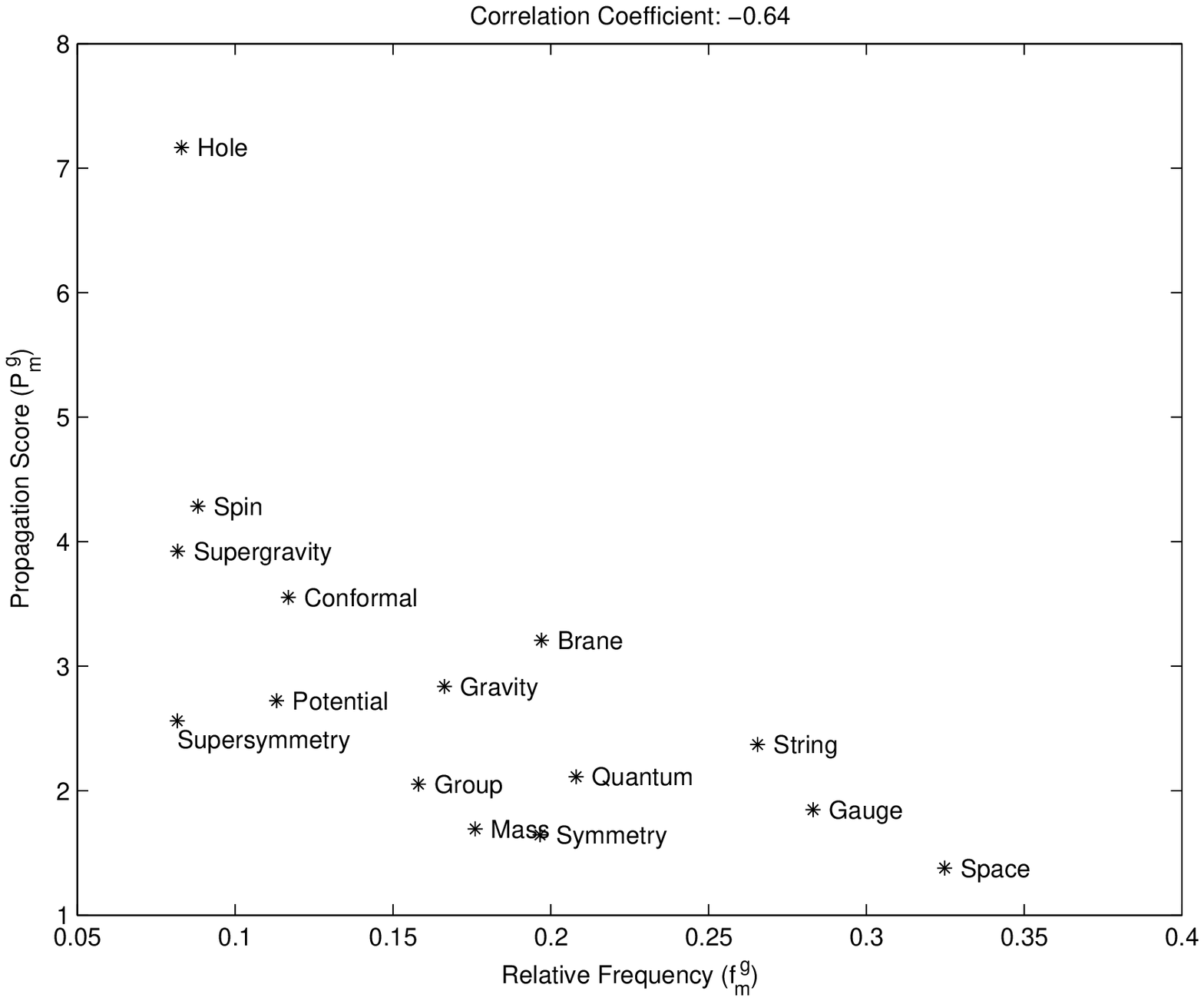,width=9truecm}
\caption{The relative frequency ($f_m^{g}$) and propagation score ($P_m^{g}$) of gendered papers plotted against each other.}
\end{center}
\end{figure}

Figure 6 shows a scatter plot where the coordinates of each 15 meme is given by its relative frequency ($f_m^{g}$) and propagation score ($P_m^{g}$). When the relative frequency and propagation score are plotted against each other, our results are in line with the outcomes presented in reference \cite{Kuhn}, showing that less frequent memes tend to propagate more (via citation links).
A possible reason for such a simple relation between the relative frequencies and propagation scores of scientific memes may rely on the fact that the less frequent ones are presumably more informative and therefore occur less often. Likewise, functional words - such as determiners and prepositions - carrying less meaning, occur very frequently and therefore occupy the most central positions in linguistic (co-occurrence) networks (\cite{Kur}, \cite{Ara16}).

Computing the correlation coefficient between the values of $f_m^{g}$ and $P_m^{g}$ for the set of gendered papers yields $-0.64$. As the propagation score of a meme captures how interesting it is for the scientific community, our results confirm that being interesting is inversely related to occurring frequently. The scatter plot in Figure 6 shows that such a simple relation holds when citation ties are gendered.

Not surprisingly and given that information on the gender of the authors that one cites is usually missing, the transmission of memes are free of gender-homophily trends in citation choices.

There is a broad literature (\cite{Grano}, \cite{Sorenson}, \cite{Kri},  \cite{Borgatti}, \cite{Cainelli}) on social relations (social networks included) showing that many social systems create contexts in which homophilic relationships hold. From friendship, co-membership and marriage, several studies have discussed the role the similarity plays in the creation of human relationships. The phenomena of  establishing ties with similar individuals have been extensively studied through network approaches, regardless whether similarity is based on age, religion, education, occupation, or gender. Recent research on the structure of citation networks \cite{Ciotti} presents a method for measuring the similarity between articles through the overlap between the bibliographic lists of references included in these articles (cited papers). One related study is discussed by Ramon Ferrer i Cancho \cite{Ferrer} with the definition of a similarity network between articles on linguistic, cognitive and brain networks. There, instead of bibliographies, the similarity between articles is measured on the basis of similar words used in the abstract of the articles. Therefore, the network approach allows for clustering articles on linguistic networks into different modules depending on whether they deal with semantics or functional brain networks.

When the gender aspect is considered, a large scale analysis on gendered authorship \cite{West} based on eight million papers across multiple areas reveals that women are significantly under-represented as authors of single-authored papers. Ara\'{u}jo and Fontainha arrived to close results when analyzing gender authorship of scientific papers through a network approach \cite{Ara17}. This paper, seeking to build upon the previous literature on gender aspects in research transmission adds to usual citation analysis the memes approach and propagation score methodology. Our computation of the propagation scores of memes characterized by the gendered authorship of the citing and cited papers allows for investigating the combined effect of meme inheritance and gendered transmission.

In so doing and despite the small difference accounted for the meme "Spin", our results show that the propagation of the selected  memes  does not seem to be influenced by the gendered authorship. The selected memes do not spread differently from either male or female cited authors. Neither female or male inheritance seems to favor the propagation of any of the selected memes. Likewise, with a single exception, the memes that we analyzed were not found to propagate more easily via male or female inheritance.

\section{Conclusion}

Our approach adds the meme inheritance notion to traditional citation analysis, as we investigate if scientific memes are inherited differently from gendered authorship. Results reveal that the inheritance process does not differ by gender.
The descriptive analysis suggest the absence of any gender-homophily trend in citation ties.
The empirical analysis also show that there is a very unbalanced scientific output by gender in the scientific domain under analysis. Women represent  about $\frac{1}{10}$ of the authorship outputs. Moreover, our results are in line with the results presented in reference \cite{Kuhn}, confirming that there is a simple relation between the frequency of occurrence of a scientific meme and its propagation score via citation links. Here we show that such a simple relation also holds when citation ties are gendered.

The paper contributes to providing a more precise characterization of women in research, and in doing so, it can contribute to informing the design, follow-up and evaluation of research programs and projects that include gender balance in their objectives. The EU Research and Innovation programme, Horizon 2020, for example, specifically stipulates three objectives pertaining to gender equality: to foster gender balance in Horizon 2020 research teams; to ensure gender balance in decision-making; and to integrate gender/sex analysis in research and innovation (R\&I) content (eige.europa).
The present paper, contributing to a better understanding of knowledge transmission by gender will also help to increase the quality and relevance of the R\&I outputs production and diffusion processes (\cite{European}).

Concerning citation analysis and sciencitometrics, this paper goes a step further investigating the interplay of memes transmission and gendered authorship. The methodology can be useful for academics conducting citation studies and knowledge diffusion analyses. For big data developers, owners, editors, administrators, and funding agencies, the present study also enlarges the horizons of knowledge production and dissemination. In particular, not only are the owners of big databases in a strategic position, but they also have the resources to develop new tools to deal with the lack of information on gender.
In the future, when the big bibliometric databases start to include it as a regular procedure, this study can be replicated on a broader scope, free of missing data.

Future research work is planned to further approach citation networks of gendered authors. Following the work of Ciotti and co-authors \cite{Ciotti} we envision the application of our gender-oriented perspective to the definition of networks of authors based on the overlap between their common references. Therefore, the network approach might allow for clustering gendered authors into different groups depending on multiple characteristics of their bibliographic references. Moreover, applying well-known statistical tools inspired by network studies in other domains, may bring important contributions to the study of  networks of scientific collaboration. We envision that, the finding of structural differences between citation networks of different types
may be indicative of their usefulness in a more applied context as tools for knowledge diffusion and transfer.

\bigskip

\textbf{Acknowledgement}

\bigskip

Financial support by FCT (Funda\c{c}\~{a}o
para a Ci\^{e}ncia e a Tecnologia), Portugal is gratefully acknowledged.
This article is part of the Strategic Project: UID/ECO/00436/2013. The authors thank R. Vilela Mendes for providing help in the identification of important physics concepts.
The research reported in this paper is based on the findings of the PLOTINA project ("Promoting gender balance and inclusion in research, innovation and training"), which has received funding from the  European Union's Horizon 2020 research and innovation programme, under Grant Agreement N. 666008 (www.plotina.eu). The views and opinions expressed in this publication are the sole responsibility of the authors and do not necessarily reflect the views of the European Commission.

\end{document}